\documentstyle[pra, aps, twocolumn, graphics, epsfig,
floats]{revtex}
\newcommand{\ket}[1]{\left | \, #1 \right \rangle}

\begin{document}
\title{Violations of Local Realism in the Innsbruck GHZ experiment}
\author{Marek \.Zukowski\\
{\protect\small\sl Instytut Fizyki Teoretycznej i Astrofizyki,}\\
{\protect\small\sl Uniwersytet Gda\'nski, PL-80-952 Gda\'nsk, Poland;}\\
{\protect\small\sl Institut f\"ur Experimentalphysik,}\\
{\protect\small\sl Universit\"at
Innsbruck, A-6020 Innsbruck, Austria}\\ }
\date{\today}
\maketitle

\begin{abstract}
It is shown that a careful analysis of the `wrong' events (those not
present in the usual formulation of the GHZ argument),
which are a necessary feature of the tests of local realism involving
independent sources, permits one to show that there is no local
realistic model, which is capable to describe recent GHZ  experiments
performed in Innsbruck.
\end{abstract}
\pacs{3.65bz}



The purely quantum concept of entanglement 
is behind the bizarre
quantum predictions for multiparticle tests of local realism. 
 The discovery of the Bell theorem without
inequalities \cite{GHZ89}, has led to several experimental proposals
to test the very existence of maximally entangled three particle
states. In 1998, an experiment of this type,
involving the  technique  of entangling
emissions from independent sources  \cite{ZZHE} has been performed in
Innsbruck \cite{INNSBRUCK}. The expected GHZ type correlations
were observed. 

Bell-GHZ experiments involving independent sources of particles, even in 
their idealized versions,
have one characteristic trait. Some observed events do not
follow the usual pattern
of the GHZ correlations. 
The aim of this paper is to show, following the ideas of ref \cite{POP}, 
that those `wrong' events are
irrelevant in the derivation of a GHZ-type contradiction for the 
quantum predictions for {\it entire} pattern of events  in experimental
configuration of ref. \cite{INNSBRUCK}. Thus, in principle, 
the Innsbruck experiment
could be a GHZ-test of the basic premises of local realism, and the only
obstacles are of a technical nature. {\it Nota bene},  
such is the status 
of all Bell-tests carried out thus far
(recall, e.g., the quantum efficiency loophole).
 
It will be  shown that  
in any local realistic theory  those wrong events  must happen (or not)
irrespective of what observables are chosen to be measured by the
remote observers at the three spatially separated stations.
Therefore, the full subset of
the ``right" events must also be independent of those choices. 
Thus, the GHZ argumentation can be confined
to only
the ``right" events.


The technique for observation of GHZ effects employed in Innsbruck
 is a  development of the
methods that have been used in experiments on
quantum teleportation and entanglement swapping
\cite{BOU97}. The GHZ experiment relies upon a
spontaneous generation
of two pairs of polarization entangled photons
 during the passage of a single laser pump pulse through
a type II parametric down conversion crystal.
To
transform the two pairs of polarization entangled photons into three
entangled photons, via  the collapse of
the wave vector caused by the registration of the fourth photon,
the detection was  performed under 
conditions
which make it absolutely impossible
to distinguish from which pair the
registered trigger photon originated \cite{ZEIL97}.

The 
crystal was pumped by a short (200 fs) pulses of UV laser light. 
The efficiency for the pulse to create a
single pair was of the order of $10^{-4}$, what clearly
justifies a perturbation approach in quantum mechanical 
calculations. The pair
creation was occurring under the specific version of type II phase
matching condition \cite{BBO}: when photon in beam $a$ (fig.1) was
horizontally ($H$) polarized and the photon in beam $b$ was vertically
polarized ($V$), or other way round. The polarization part of the
state can be written as $ \label{polstate}
\frac{1}{\sqrt{2}}(\ket{H}_a\ket{V}_b - \ket{V}_a\ket{H}_b)\,
$
 where, e.g., $\ket{H}_a$ describes a horizontally polarized photon
in beam $a$.

To describe the quantum process the usual approximations can be made.
One can assume perfect phase matching, thus only two phase matched
directions can be singled out.
Further, we can demand that the idler and signal photon
frequencies satisfy perfect energy conservation conditions
with the pump photons, which is
described by the $\delta(\omega-\omega_i-\omega_s)$ factor
(where $\omega$ is one of the frequencies in the frequency band of
the pulse, and $\omega_i$ and $\omega_s$  are the frequencies of
the idler and signal photons).
The pump pulse can be described as a classical wavepacket (no-depletion)
with one single direction for all wave vectors.
Finally, 
the square nonlinearity, $\chi^{(2)}$, of the electric polarizability
can be assumed constant within the considered frequency ranges.
Thus, the interaction Hamiltonian responsible
for such a process can be given the following approximate form

\begin{eqnarray}
&H(t) = k\chi^{(2)}\int d\omega\int d\omega_a\int d\omega_b V(\omega)
e^{i\omega t}
\delta(\omega-\omega_a-\omega_b)& \nonumber\\
&
\left(a^{\dagger}_V(\omega_a)b^{\dagger}_H
(\omega_b) - a^{\dagger}_H(\omega_a)b^{\dagger}_V
(\omega_b) + h.c.
\right). & \nonumber\\
\end{eqnarray}
Here $k$ is the effective coupling constant, the frequency
profile of the pulse is denoted by $V(\omega)$,
$h.c.$
stands for the Hermitian conjugate of an earlier expression,  and
the convention used for the creation operators is that they are denoted
by the same letter as the spatial propagation mode which they
represent.
After performing the integration over $\omega$ and passing to
the interaction picture, the Hamiltonian simplifies to
 \begin{eqnarray}
&H_I = k\chi^{(2)}\int d\omega_a\int d\omega_b V(\omega_a + \omega_b)
& \nonumber\\
&
\left(a^{\dagger}_V(\omega_a)b^{\dagger}_H
(\omega_b) - a^{\dagger}_H(\omega_a)b^{\dagger}_V
(\omega_b) + h.c.
\right), & \nonumber\\
\end{eqnarray}
i.e., it is independent of time (the photon operators are in the Schr\"odinger
picture). If one assumes a finite interaction time
$T=L/c$, where $L$ is the length of the crystal (this is
equivalent to assuming that the crystal is longer than the pulse, however
this assumption is not essential to get the final conclusion of the analysis)
the second order perturbation
term is given by:
\begin{equation}
\frac{1}{2}(TH_I/i\hbar )^2\ket{vacuum}.
\label{x}
\end{equation}
The non-vacuum part of (\ref{x}) leads to
spontaneous creation of two {\it identical} pairs of photons, and the
entanglement in frequency and polarization exists only
between photons of each of the pairs.

The almost perfect temporal correlation of the photons of one pair (due to the
entanglement in the frequencies) must be irreversibly
destroyed, since it can reveal which two photons come from one
entangled pair (see \cite{ZZHE}, \cite{ZZW},
\cite{ZEIL97}).
To this end,
the photons had to pass through two
identical filters of frequency passband  narrower than the
frequency spread of the pump pulse. If a photon emerges from the
filter, its coherence time is longer than the temporal pump pulse
width, and thus the original
tight temporal correlation is lost.

The experimental setup (fig.1) was
such that beam $b$ continues towards a 50/50 polarization-independent
beamsplitter, and beam $a$ towards a polarizing beamsplitter, which
transmits the $H$ photons towards the trigger detector T and reflects
the $V$ photons. From each beamsplitter one output is directed to a
second polarizing beamsplitter, which likewise transmits $H$
polarization and reflects $V$ polarization. A half-wave plate at an angle of
$22.5^{\circ}$  rotates the $V$ polarization of the photons
reflected by the first polarizing beamsplitter into a $45^{\circ}$
polarization. One has $\ket{45^{\circ}}=
\frac{1}{\sqrt{2}}(\ket{V}+\ket{H})$.

For simplicity, we
shall for the moment forget about the frequency entanglement (and tacitly
assume that all four photons have passed the filters, compare fig.
1), and the two pair emission by

\begin{equation}
\gamma^2\left(a^{\dagger}_Vb^{\dagger}_H
 - a^{\dagger}_Hb^{\dagger}_V
\right)^2\ket{vacuum}
\end{equation}
i.e. we prune the frequency from the formulas
($\gamma$ is a new coupling
strength factor, which also may contain the phase of the pump beam, however
this phase plays no role in the sequel).

The action of the polarization independent beamsplitter can be
described by the following relation between the input mode
annihilation operator $b_X$ (where $X=V$ or $H$), and the two outputs
(beams $c$ and $g$): $ b_X=\frac{1}{\sqrt{2}}(c_X + g_X).$ As the
polarizing beamsplitter in the beam $a$ simply distributes the photons
of appropriate polarization to two separate beams, we shall denote the
annihilation operator of the mode in the transmission beam by $a_H$,
and in the reflected beam by $a_V$. This $a_V$ mode is transformed
into $a_{45}$ by the half-wave plate. The second
polarizing beamsplitter again reflects the $V$ polarization and
transmits the $H$ one (it acts in the same way upon both input beams).
Therefore its action upon the relevant annihilation operators can be
described by $ a_{45}=\sqrt{1/2}(h_H+z_V),$ and $ c_V=h_V,$ and  finally $
c_H=z_H,$ where $h$ and $z$ are the two output beams of the polarizing
beamsplitter (fig.1).

The  triggering event for the observation process  is a registration
of a {\it single} photon by a detector located in the $a_H$
output of the first polarizing beamsplitter. If two photons enter
 the trigger detector,
we are left with only two other photons - no GHZ effect is possible.
In principle  it is possible to  discriminate  operationally two and
single photon events \cite{DOUBLECLICK}), and we shall assume that
this is the case in our idealized situation. Therefore from the
initial state one can drop the term proportional to $a^{\dagger 2}_H$,
as it always leads to a wrong type of the triggering event, and the
term proportional to $a^{\dagger 2}_V$ as it cannot produce any click
at the trigger detector. One is left only with
\begin{equation}
2\gamma^2a^{\dagger}_Ha^{\dagger}_Vb^{\dagger}_Vb^{\dagger}_H
\ket{vacuum},
\end{equation}
which under the action of the optical devices described above
transforms into
\begin{equation}
\frac{1}{\sqrt{2}}\gamma^2a^{\dagger}_H(z^{\dagger}_V+h^{\dagger}_H)
(h^{\dagger}_V+g^{\dagger}_V)(g^{\dagger}_H+z^{\dagger}_H)
\ket{vacuum}.
\label{fin1}
\end{equation}
The only term in (\ref{fin1}) which has
the property that in each beam one has
one and only one photon is
\begin{equation}
\frac{1}{\sqrt{2}}\gamma^2a^{\dagger}_H
\left(
g^{\dagger}_Hh^{\dagger}_Vz^{\dagger}_V+
g^{\dagger}_Vh^{\dagger}_Hz^{\dagger}_H\right)
\ket{vacuum}.
\label{GHZ}
\end{equation}
It is  responsible for the GHZ polarization
correlations observed in the experiment.


Imagine that three spatially separated observers, say Danny, Mike and Anton,
 can detect any polarization state of the photon
in, respectively, the beams $g$,
 $h$, and  in $z$.
For brevity, whatever polarization one of them
decides to measure, we say that he chooses a certain `orientation
of his polarizer'. Also we shall assume that their perfectly efficient 
detectors
distinguish between
counts due to  two photons and those caused by only one photon.
They set orientations of
their polarizers at random, selecting either observation of left and right
circular polarizations, or linear polarizations at $45^o$ and $-45^o$.
If the state of the photons were fully described by (\ref{GHZ}), then
such measurements would suffice to formulate
the GHZ paradox \cite{MERMIN}, \cite{INNSBRUCK}. 
However, the state (\ref{fin1}) contains also other terms, 
except (\ref{GHZ}), which are
responsible for ``wrong" events (i.e., events not present in the
original GHZ argument for three particle polarization correlations),
namely
\begin{eqnarray}
\frac{1}{\sqrt{2}}\gamma^2a^{\dagger}_H
\left(
g^{\dagger}_Hg^{\dagger}_Vz^{\dagger}_V+
g^{\dagger}_Hh^{\dagger}_Hh^{\dagger}_V+
h^{\dagger}_Hg^{\dagger}_Hg^{\dagger}_V+\right.\nonumber\\
\left.
z^{\dagger}_Hz^{\dagger}_Vh^{\dagger}_V+
z^{\dagger}_Hz^{\dagger}_Vg^{\dagger}_V+
z^{\dagger}_Hh^{\dagger}_Hh^{\dagger}_V
\right)
\ket{vacuum}.
\label{fin}
\end{eqnarray}

The wrong events have the following
trait: {\it they come in pairs} - there must be one two-photon event at a
certain observation station, necessarily accompanied by a one
non-detection event at one another station. Note, that a non-detection
event is in principle observable, as we have at our disposal the
trigger event, only after which we can expect, at well defined moments
of time, to observe detections at the three remote stations.

The other type of wrong events is due to the first order process,
in which only one pair is created, i.e. the following initial state
is responsible for such events
\begin{equation}
\gamma\left(a^{\dagger}_Vb^{\dagger}_H
- a^{\dagger}_Hb^{\dagger}_V
\right)\ket{vacuum}.
\end{equation}
Now, if the trigger fires, then there can be recorded only one photon
in one of the beam $h$ (Mike) or $z$ (Anton). Danny 
would not register anything in $g$.
I.e., we shall always have {\it two} non-detection events.

There is one more possibility of additional wrong events to occur. 
The action of a filter is
essentially to remove photons from a beam. If our filter in beam $a$
removes two $H$ polarized photons, there is no problem: the trigger
event would not occur. If it removes one $H$ polarized photon, then
either we again shall not have the triggering event, or suddenly one
of the neglected terms in the two-pair emission, namely
\begin{equation}
\gamma^2 (a^{\dagger}_Hb^{\dagger}_V)^2
\ket{vacuum},
\end{equation}
leads to a (seemingly)  right triggering event, and some additional
``wrong" events at the three stations. Also the removal of one of two
photons by
 the filter in beam $b$ can lead to unwanted effects (missing counts).

In order to avoid those additional wrong events, one should redefine the
trigger event. Removal of a photon from a beam by a filter is in principle
recordable \cite{ANTON}. Just imagine as a toy model filter a prism,
behind
 which
we select only photons which deviate into a narrow band of angles, and
we detect all photons that deviate differently.
One can place the  filters, and the trigger detector,  close to the PDC
crystal and redefine the
triggering event to be: a detection of a 
single photon at $T$,  {\it and} no click at the detecting
devices registering the photons  that failed to pass the filters.


The most important trait of the remaining wrong 
events is that {\it they always come in pairs}. The local realistic 
theory of the experiment, must therefore
have very specific traits. For clarity,  a local
hidden variable model of such a theory will be used. In such a model
the local events (i.e. events at one of the observation stations) are
determined by the value of the hidden variables describing the
experiment, usually denoted by the symbol $\lambda$, and the local
macroscopic controllable parameter set by the local observer (here the
orientations of the polarizers set by Danny, Mike and Anton; they will
be symbolically denoted by
$\theta_g$, $\theta_h$ and $\theta_z$).

 Danny, Mike and Anton, when the {\it right} event
occurs at their detection station, read out the value of dichotomic
variable. If one of them decides to measure the circular
polarizations, then he can say that the result is $+1$, if it is right
handed and $-1$, if it is left handed. The same procedure applies to
the other choice (registration of $45^o$ polarized photon is $+1$,
etc.). Let us also ascribe the value $0$ to any locally observed
wrong event. One can denote the overall results, and their dependence
on the hidden variable and the orientation of the {\it local}
polarizer by respectively $G(\theta_g, \lambda)$, $H(\theta_h,
\lambda)$ and $Z(\theta_z, \lambda)$ (all these functions
have the set of values composed of solely $1$, $-1$ and $0$). The
distribution of the hidden variables responsible for the overall
statistics of the experiment can be denoted as usually by
$\rho(\lambda)$.

Let us introduce the following function
\begin{equation}
\Sigma(\theta_g,\theta_h, \theta_z; \lambda)=
|G(\theta_g, \lambda)|+|H(\theta_h,
\lambda)|+|Z(\theta_z, \lambda)|.
\end{equation}
For ``right" events at all stations 
$\Sigma(\theta_g,\theta_h, \theta_z; \lambda)=3$.
If there is at least one ``wrong" event then 
$\Sigma(\theta_g,\theta_h, \theta_z; \lambda)=1$ (since there is 
always a second wrong event!). One never has $\Sigma$ equal to $0$ or $2$.

Imagine now that the occurrence of a local wrong event may depend not only 
on the value of the hidden variable $\lambda$, but also on the orientation 
of the local polarizer. E.g. for $\theta_z\neq\theta_z'$, and a 
fixed $\lambda$ at Anton's station
one may have, for the same $\lambda$,
  $|Z(\theta_z, \lambda)|=1$ and  $|Z(\theta_z', \lambda)|=0$. 
Therefore, since  $\Sigma(\theta_g,\theta_h, \theta_z; \lambda)=3$ or $1$, 
then it follows that $\Sigma(\theta_g,\theta_h, \theta_z'; \lambda)=2$ or $0$
which is impossible! We arrive at similar contradictions for the other 
possible cases. Therefore one concludes that 
the modulus of the local result is independent of the orientation of the local 
polarizer,
or in other words `wrong' events are predetermined by solely the hidden
variables.     
The product of these moduli,
$\chi(\theta_g, \theta_h,
\theta_z, \lambda)= |G(\theta_g, \lambda)H(\theta_h, \lambda)
Z(\theta_z, \lambda)|,$ can have only two values: $1$,
for three right events or $0$ otherwise.  
Due to the properties of the ``wrong" events, $\chi$   does not
depend upon the local polarizer orientations. Therefore, the
distribution of the hidden variables, $\rho(\lambda)$ splits into
$\rho_{right}(\lambda)=\rho(\lambda)\chi(\lambda)$ and
$\rho_{wrong}(\lambda)=\rho(\lambda)(1-\chi(\lambda))$. The correct
GHZ type events occur only for those values of $\lambda$ for which
$\rho_{right}(\lambda)\neq0$. The whole reasoning leading to the
GHZ paradox \cite{MERMIN} can therefore be limited the
$\lambda$'s belonging to this subset only.

Thus, modulo photon collection efficiency,  and linked with this
the problem of  the 
operational possibility of a correct discrimination of the
``wrong" two-photon events at a single detection station
\cite{DOUBLECLICK}, one can conclude that the Innsbruck experiment not
only indicates an underlying quantum three-particle process, due to a
GHZ entanglement, which is absolutely at odds with local realism, but
is also the very first realization of a three particle test of local
realism without inequalities. Simply, since the three particle interference
fringes are too deep for any local hidden variable theory to explain
the entire pattern of events
\cite{KASZLIKOWSKI}.

The author is very grateful to Anton Zeilinger and
Harald Weinfurter for years of discussions
concerning
the studied problems.
This work was supported by the Austrian-Polish 
Program {\it Quantum Information and Quantum Communication II} 11/98b,
and by the University of Gda\'nsk, Program BW/5400-5-0202-8.

\newpage
\begin{figure}
        \begin{center}\mbox{\input epsf \epsfxsize0.8 \columnwidth
\epsfbox{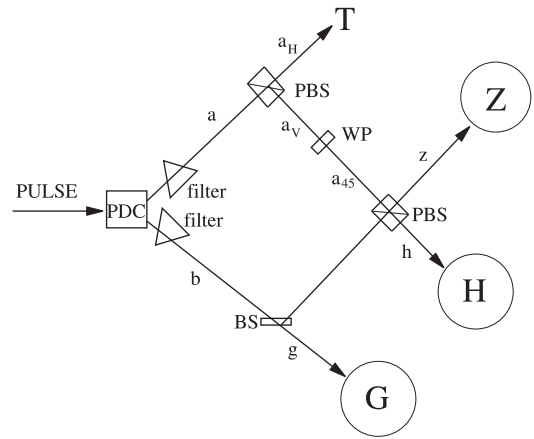}}\end{center}
        \caption{Schematic of the setup. PBS is a 
        polarizing beamsplitter, which transmits
        the $H$ polarization and reflects the $V$ polarization. BS is a 
polarization neutral ``$50-50$"
        beamsplitter. PDC stands for a parametric down conversion crystal 
(pumped by pulses originating from a laser).
        WP is a wave plate which rotates the
        $V$ polarization to $45^o$ polarization. 
        The beams in which both $H$ and $V$ polarizations propagate
        are denoted by $a$, $b$, $g$, $h$ and $z$. In the transmitted beam of 
the upper PBS
        only polarization $H$ can be present and thus it is denoted by $a_H$
        (the reflected  beam $a_V$ can carry only polarization $V$, and after 
the wave plate is denoted accordingly by
         $a_{45}$). At location $T$ a trigger detector is placed.
        At locations $G$, $H$ and $Z$ (which are widely separated) 
        three independent observers
        measure the polarization states of the incoming photons. }
       
\end{figure}


\begin{thebibliography}{99}
\bibitem{DOUBLECLICK}
To  discriminate  operationally two and single photon events,
one can e.g. split the incoming beam
into
$n$ beams and place an
ordinary detector (which cannot distinguish one- and two-photon events) in
each of these  beams. 
If $n$ is sufficiently large, the probability of two photons to enter
a single detector can be made arbitrarily small. This scheme does not
work if
the efficiency of the detectors is low. But this is  a feature of all
Bell-type experiments - the efficiency requirements are high.
\end{thebibliography}
\end{document}